\documentclass[a4paper]{article}
\usepackage{graphicx}
\usepackage{xspace}

\begin{document}

\begin{center}
{\Large \bf Calculation of light nucleus reaction cross sections in Geant4}
\end{center}

\begin{center}
{V. Uzhinsky}
\end{center}

\begin{center}
{CERN, Geneva, Switzerland and LIT, JINR, Dubna, Russia}
\end{center}

\begin{center}
(on behalf of the Geant4 hadronic group)
\end{center}

\begin{center}
\begin{minipage}{12cm}
Total reaction cross sections of light projectile nucleus ($^2{\rm H}$, $^3{\rm H}$,
$^3{\rm He}$ and $^4{\rm He}$) interactions
with nuclei are calculated using Geant4 models, and compared with experimental data. It is shown
that the models give various predictions at low energies, in the region of the Coulomb barrier.
"Shen model" (W.-Q. Shen et al., Nucl. Phys. {\bf A491} (1989) 130) is identified as an improvement
over other models.
\end{minipage}
\end{center}

Five methods/models are used in the Geant4 toolkit \cite{GEANT4}  for calculations of nucleus-nucleus cross sections:
LHEP originated from Geant3, Kox \cite{Kox}, Sihver \cite{Sihver}, Shen \cite{Shen} and Tripathi
\cite{Tripathi,TripathiLI}. They are implemented in the corresponding classes: {\it G4HadronCrossSections},
{\it G4IonsKoxCrossSection}, {\it G4IonsSihver} {\it CrossSection}, {\it G4IonsShenCrossSection},
{\it G4TripathiCrossSection}, {\it G4TripathiLightCrossSection}. The models were proposed many years ago.
Several experimental measurements of nucleus-nucleus cross sections have been published since that time.
Thus there is a new possibility to validate the models. Below a set of collected experimental data will
be presented with the model calculations performed in the Geant4 framework. The main attention is paid
to light projectile nucleus interactions.

There is also a new phenomenological model proposed by V.N. Grichine in a recent paper \cite{Grichine}.
The corresponding class is {\it G4GGNuclNuclCrossSection}.

Usually, nucleus-nucleus interactions are characterized by total, elastic and inelastic cross sections.
A total reaction cross section is the difference between the total and elastic cross sections.
The inelastic cross section can include cross sections of various reactions.  For example,
quasi-elastic scattering cross sections of processes where a projectile or a target nucleus is left
in the ground state and its collision partner becomes into an excited states, a cross section of
double quasi-elastic scattering of a process when projectile and
target nuclei are in excited states after a reaction. The excited states can be states with excitations
of discrete nuclear levels or states in the continuum region. There can be also charge-exchange reactions,
fusion reactions at low energies, multi-particle production processes at sufficiently high energies and so.

The identification of the reactions is experimentally complicated, especially when there can be
unobserved neutrons in the final states. Thus the comparison of various experimental data is not a simple
task. Quite often total reaction cross sections are estimated in the optical model
after an analysis of elastic scattering data. These estimations will be used in the comparison with model
predictions.

Theoretical models are usually phenomenological ones. The optical model, which can predict various cross
sections, is applied at low energies (below 1 GeV/u). At high energies the Glauber model
\cite{Glauber,GlauberCM,GlauberMC} is used.
In the last decades, the Glauber model with corrections has been applied for an analysis of data at
low energies. The application of the model for exotic nuclei (${}^6{\rm He}$, ${}^8{\rm He}$,
${}^{11}{\rm Li}$, ${}^{10}{\rm B}$, ${}^{16}{\rm C}$ and so on)
interactions with nuclei is an active field now.

Experimental data and Geant4 model calculations are presented for $d+A$, $t+A$,
$^3{\rm He}+A$ and $^4{\rm He}+A$ reactions in figures 1--4 . To summarize,
\begin{itemize}
\item The LHEP model gives energy independent cross sections with a sharp threshold at low energy end.
It does not provide cross sections for ${}^3{\rm He}+A$ reactions.

\item The Sihver model gives energy independent cross sections with a sharp threshold at low energy domain.
It covers all considered light projectile nuclei.

\item The Kox model gives meaningful results at all energies. The Shen model is an improved Kox model.

\item The Tripathi Light Ion model suppresses cross sections at low energies
for light target materials. It does not give results for $^3{\rm H}+A$ reactions. Its predictions start to
deviate from the experimental data for heavy targets and $^4{\rm He}$ projectile.

\end{itemize}

The Shen and Tripathi model calculations are compared with the experimental data and
with predictions of a model proposed by C.-T. Liang et al. \cite{Liang} in figures 5--8 . There is also calculations by
the Grichine's model. To summarize,
\begin{itemize}

\item The Grichine's model calculations deviate from the data in an irregular manner. It is not clear
how to improve the model.

\item The Liang et al. parameterization describes the data rather well, better than other approaches.
Its predictions deviate strongly from other model calculations at $E/A < 10$ MeV/u.
At the same time, the validity of the Liang model predictions for heavy targets is questionable.

\item The Tripathi model was tuned using mainly the optical model estimations (see Fig. 8,${}^4{\rm He}+{\rm Ca},
\ {\rm Ni}$ cross sections). In general, the estimations are below experimental data.
\end{itemize}

As seen, all models give different predictions at low energies. It is unknown if any of the models are
reliable. All the models are available in the Geant4 toolkit, and can be used.

Two models, Shen and Tripathi, should be considered for a future development.
Theoretical estimations of the cross sections at low and high energies would be useful for this development.

Because the upper limit of the Tripathi model is 1 GeV/u, it should be well to use the Shen model
in various physics lists of the Geant4 in the future Geant4 releases. The Tripathi model may be used
for low energy applications.

The author is thankful to M. Kelsey, J. Apostolakis, V. Ivanchenko, T. Koi and A. Ribon for interest
in this work and important remarks. He is also thankful to Jose Manuel Quesada Molina for a very important
consideration of the cross section behaviour in the region of the Coulomb barrier.

\begin{figure}[cbth]
\includegraphics[width=160mm,height=200mm,clip]{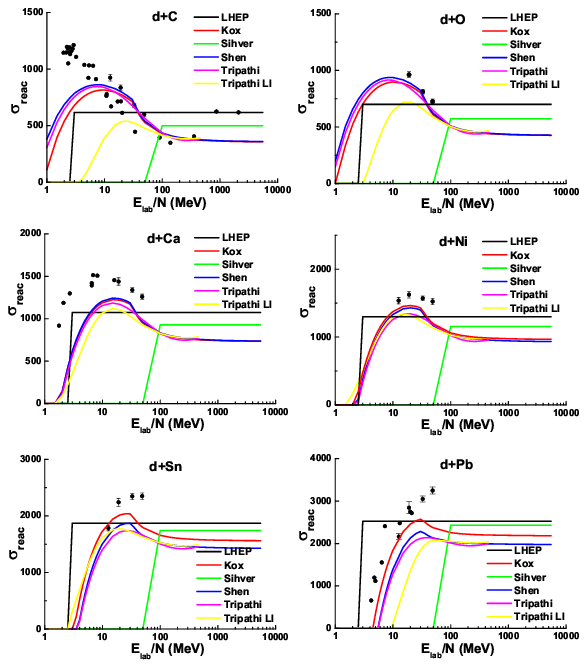}
\caption{$d+A$ total reaction cross sections. Points with error bars are experimental data
\protect{\cite{MayoNP62,AucePRC53,JarosPRC18}}. Points without error bars are
optical model estimations \protect{\cite{DeVriesPRC22}}.
}
\label{Fig1}
\end{figure}

\begin{figure}[cbth]
\includegraphics[width=160mm,height=200mm,clip]{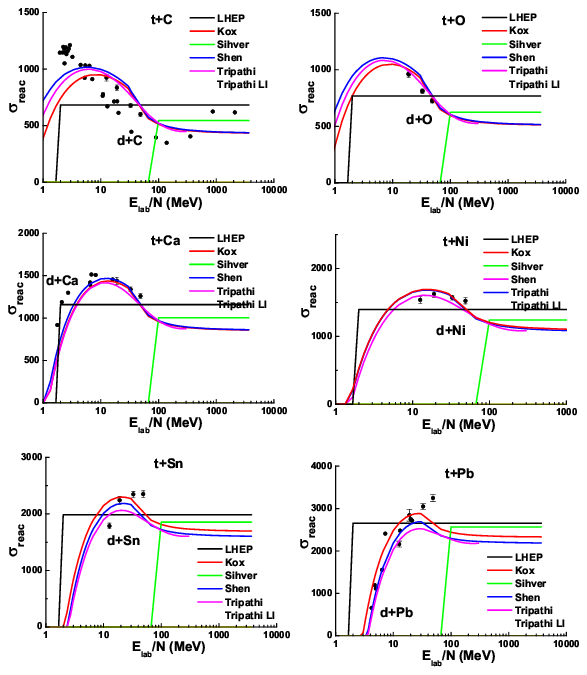}
\caption{
$t+A$ total reaction cross sections. Points are experimental data and optical model estimations
for $d+A$ reactions (see Fig. 1).
}
\label{Fig2}
\end{figure}

\begin{figure}[cbth]
\includegraphics[width=160mm,height=200mm,clip]{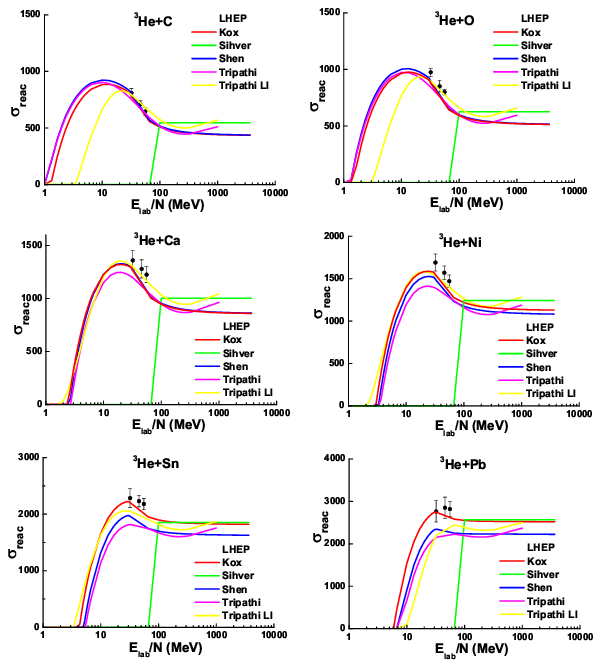}
\caption{
$^3{\rm He}+A$ total reaction cross sections. Points are experimental data \protect{\cite{IngeNPA696}}.
}
\label{Fig3}
\end{figure}

\begin{figure}[cbth]
\includegraphics[width=160mm,height=200mm,clip]{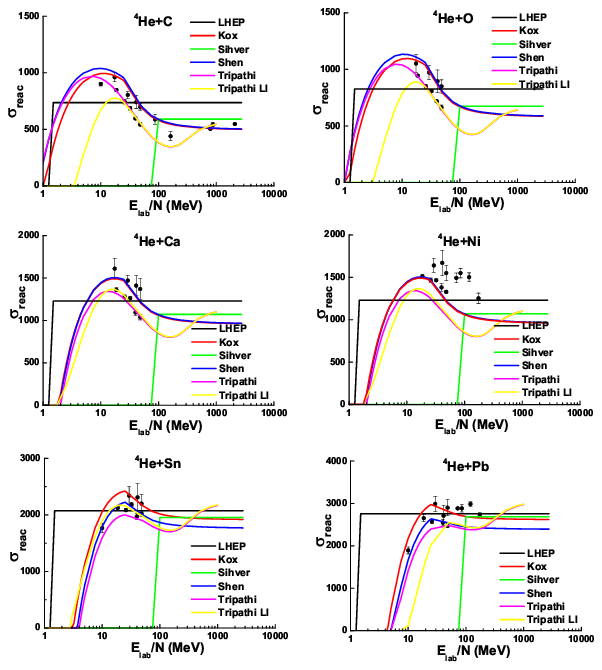}
\caption{
$^4{\rm He}+A$ total reaction cross sections. Points with error bars are experimental data
\protect{\cite{IgoPR131,DeVriesPRC26,AucePRC50,JarosPRC18,IngeNPA676,BoninNPA445}}. Points without error bars are
optical model estimations \protect{\cite{DeVriesPRC22}}.
}
\label{Fig4}
\end{figure}

\begin{figure}[cbth]
\includegraphics[width=160mm,height=200mm,clip]{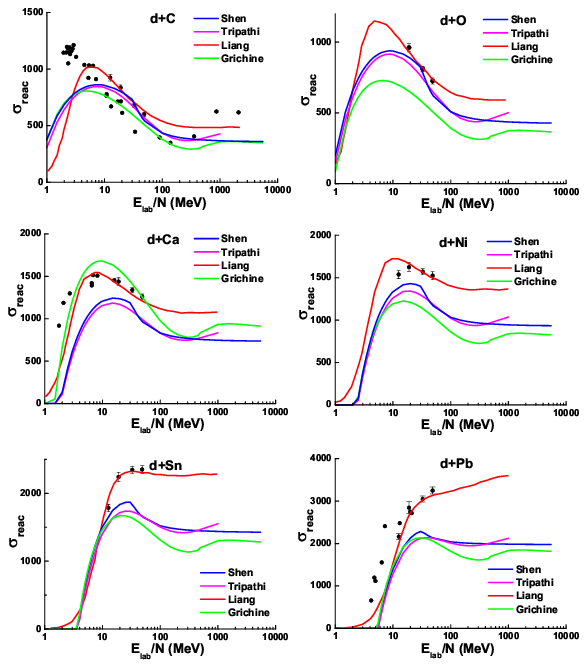}
\caption{
$d+A$ total reaction cross sections. Points with error bars are experimental data
\protect{\cite{MayoNP62,AucePRC53,JarosPRC18}}. Points without error bars are
optical model estimations \protect{\cite{DeVriesPRC22}}.
}
\label{Fig5}
\end{figure}

\begin{figure}[cbth]
\includegraphics[width=160mm,height=200mm,clip]{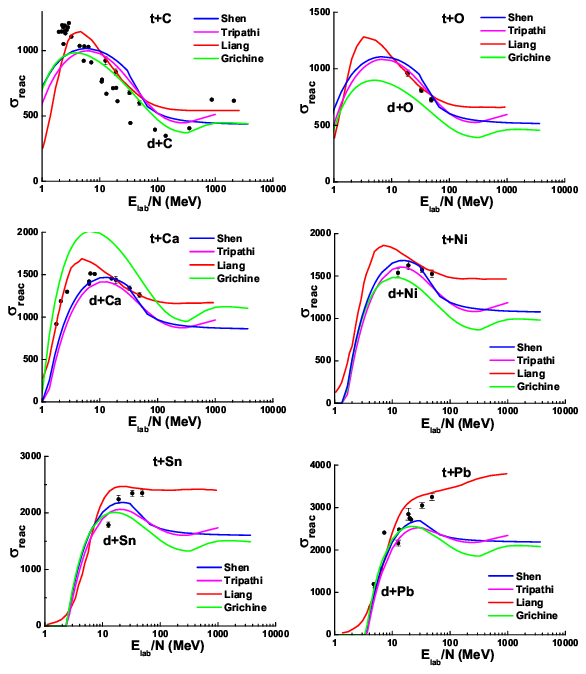}
\caption{
$t+A$ total reaction cross sections. Points are experimental data and optical model estimations
for $d+A$ reactions (see Fig. 1).
}
\label{Fig6}
\end{figure}

\begin{figure}[cbth]
\includegraphics[width=160mm,height=200mm,clip]{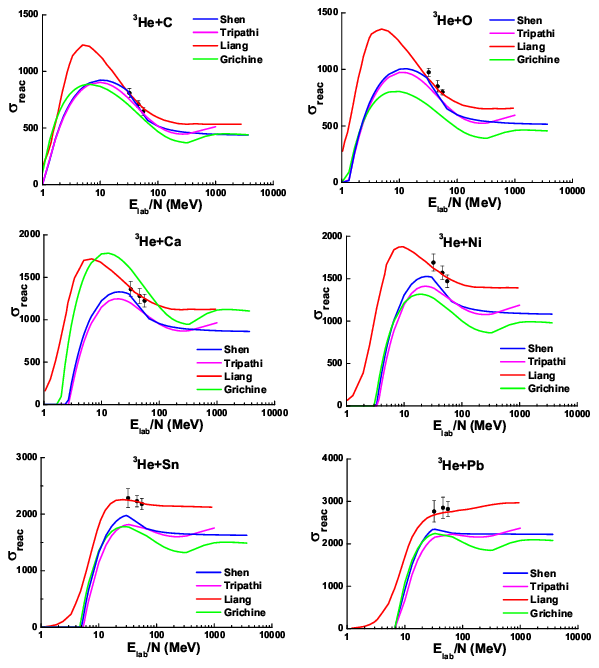}
\caption{
$^3{\rm He}+A$ total reaction cross sections. Points are experimental data \protect{\cite{IngeNPA696}}.
}
\label{Fig7}
\end{figure}

\begin{figure}[cbth]
\includegraphics[width=160mm,height=200mm,clip]{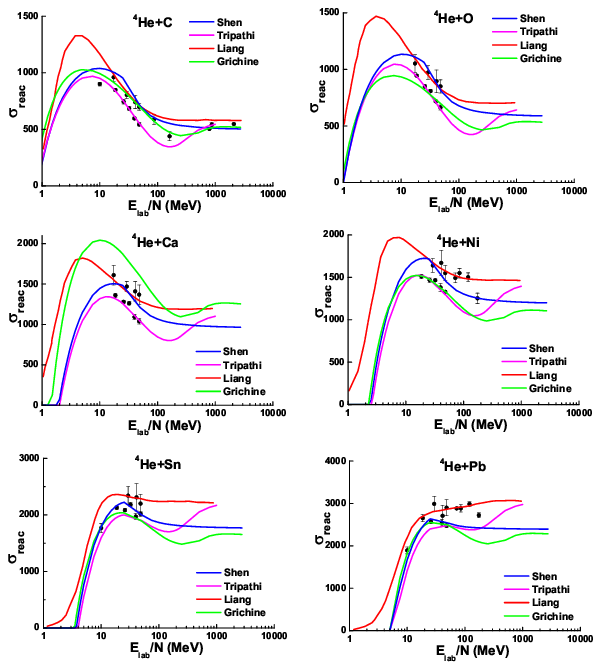}
\caption{
$^4{\rm He}+A$ total reaction cross sections. Points with error bars are experimental data
\protect{\cite{IgoPR131,DeVriesPRC26,AucePRC50,JarosPRC18,IngeNPA676,BoninNPA445}}. Points without error bars are
optical model estimations \protect{\cite{DeVriesPRC22}}.
}
\label{Fig8}
\end{figure}

\end{document}